\documentclass{article}
\usepackage{graphicx}
\usepackage[english]{babel}

\newcommand{\bk}{{\bf{k}}}
\newcommand{\bl}{{\bf{l}}}

\newcommand{\bq}{{\bf{q}}}

\newcommand{\bkappa}{\mbox{\boldmath $\kappa$}}

\newcommand{\aem}{\mbox{$\alpha_{\rm{em}}$}}
\title{
\includegraphics[width=0.35\textwidth]{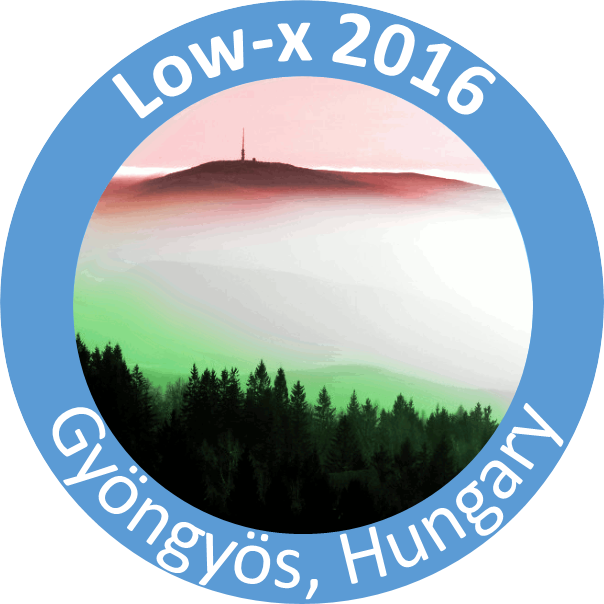}\\[1cm]
Drell-Yan production at forward rapidities: a hybrid factorization approach
}
\author{{Wolfgang Sch\"afer$^1$, Antoni Szczurek$^{1,2}$}\\[1ex]
$^1$Institute of Nuclear Physics, Polish Academy of Sciences\\ Krak\'ow, Poland\\
$^2$University of Rzesz\'ow, Rzesz\'ow, Poland \\
}

\begin{document}

\fontfamily{lmss}\selectfont
\maketitle

\begin{abstract}
We discuss the Drell-Yan production of dileptons at high energies 
in the forward rapidity region of proton-proton collisions in a hybrid high-energy approach.
This approach uses unintegrated gluon distributions in one proton 
and collinear quark/antiquark distributions in the second proton.

We compute various distributions for
the case of low-mass dilepton production and compare 
to the LHCb and ATLAS experimental data on dilepton mass distributions.
In distinction to dipole approaches, we include four
Drell-Yan structure functions as well as cuts at the level of lepton kinematics. 
The impact of the interference
structure functions is rather small for typical experimental
cuts. We find that both side contributions ($g q/\bar q$ and $q/\bar q g$) 
have to be included even for the LHCb rapidity coverage which
is in contradiction with what is usually done in the dipole approach.
We present results for different unintegrated gluon distributions from
the literature. Some of them include saturation effects, but we
see no clear hints of saturation even at small $M_{ll}$.
\end{abstract}

\section{Introduction}
Drell-Yan production in the forward direction is dominated by the quark-gluon 
fusion, where especially at not too large invariant masses of the dilepton
system the gluon density is probed at low values of the 
longitudinal momentum fraction $x$. One might therefore probe a kinematic range
where gluon saturation effects are potentially large.
Consequently the forward Drell-Yan process has been discussed in the
Color-Glass Condensate approach in \cite{Gelis:2002fw}.
Recently much attention has also been paid to appplications of the 
color dipole approach to the Drell-Yan process at the LHC
(see e.g. \cite{Ducati:2013cga,GolecBiernat:2010de,Basso:2015pba,Motyka:2014lya}) 
 
In this talk we instead present an alternative formulation in momentum space 
published recently in \cite{Schafer:2016qmk}.
In particular, this approach includes all the four structure functions
\cite{Oakes:1966} of the Drell-Yan process and
allows to put cuts on the momenta of 
individual leptons ($e^+ e^-$ or $\mu^+ \mu^-$). This is 
important if one wants to compare to existing experimental data.

\begin{figure}[!ht]
\includegraphics[width=.5\textwidth]{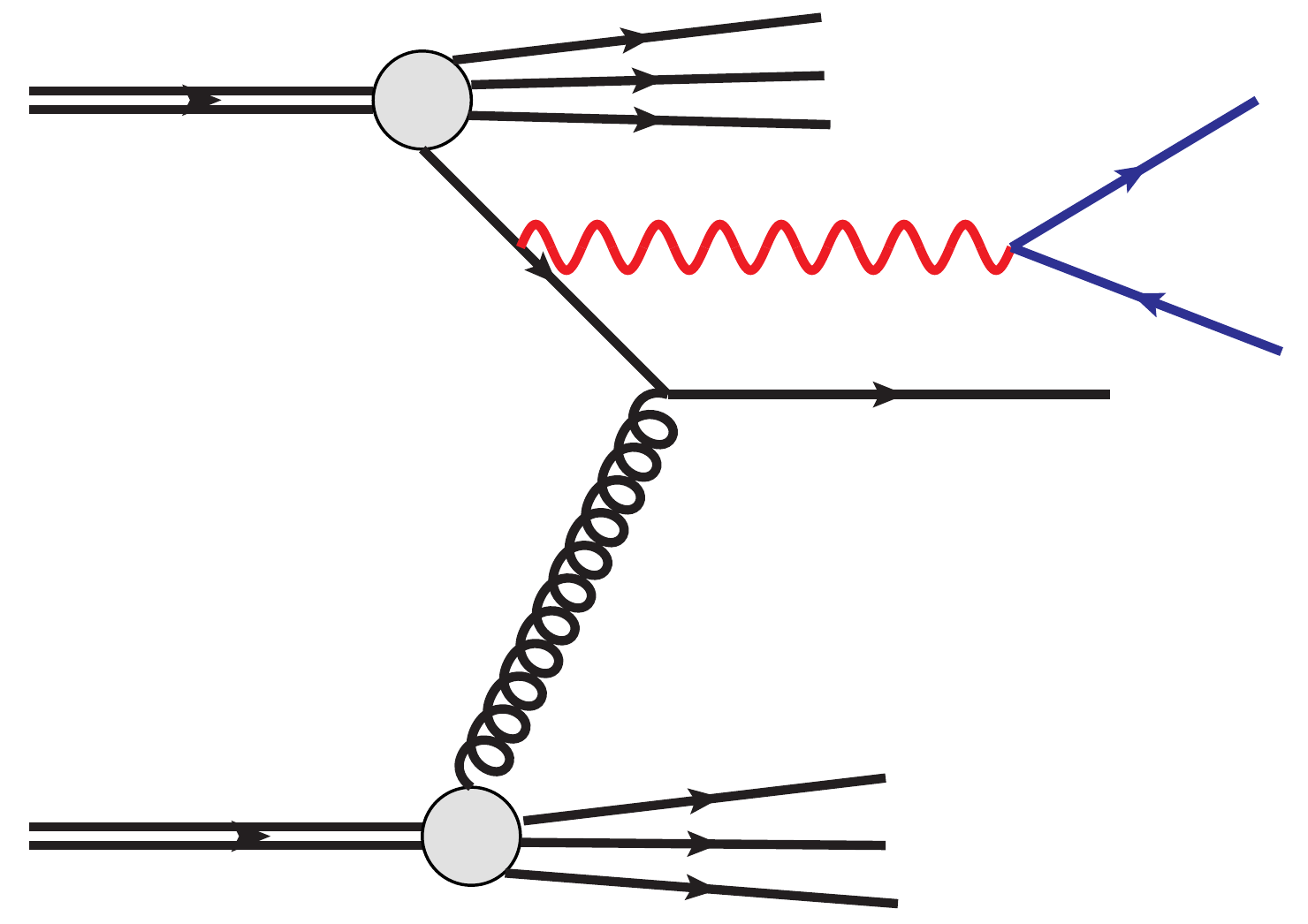}
\includegraphics[width=.5\textwidth]{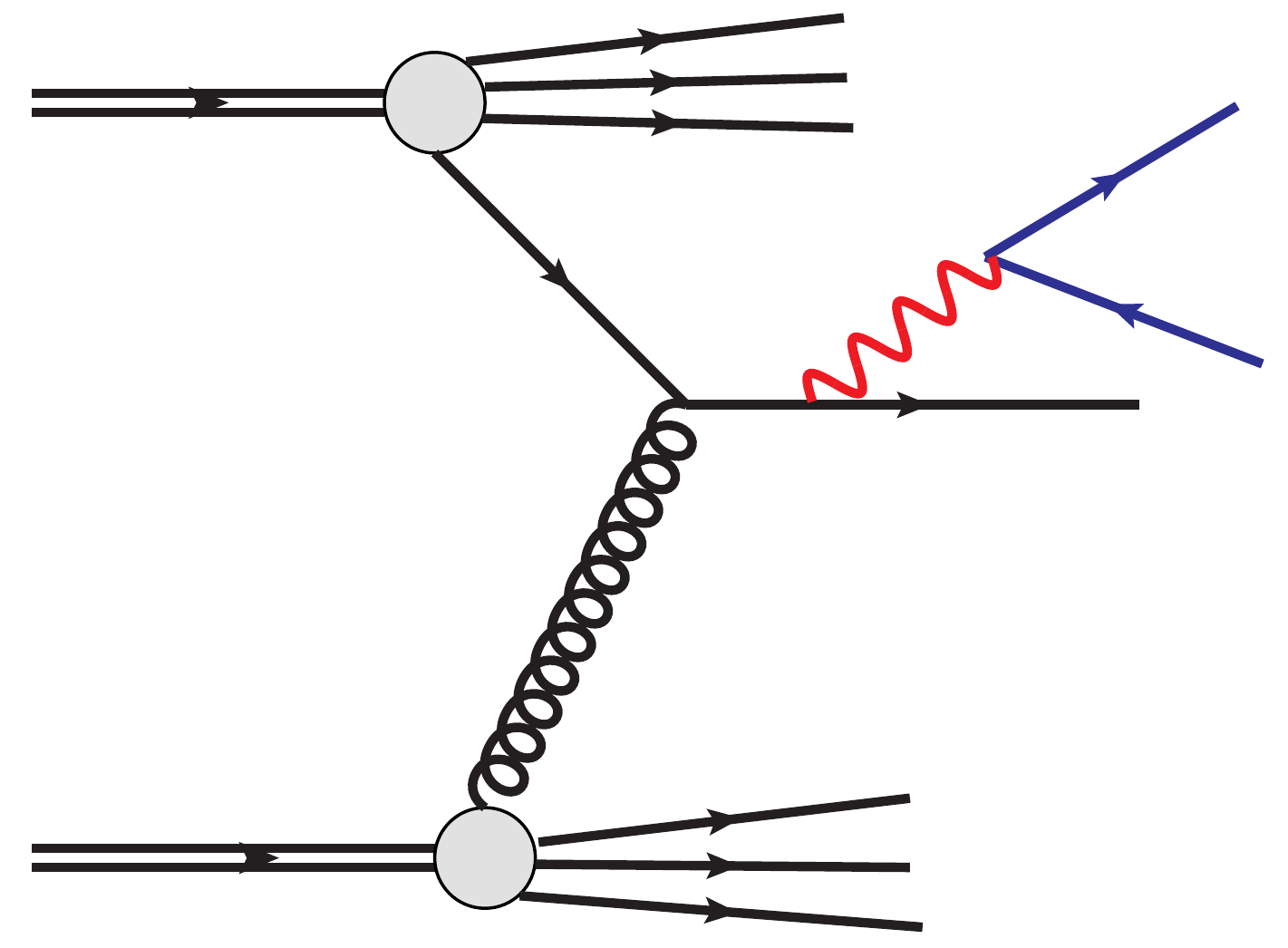}\\
\includegraphics[width=.5\textwidth]{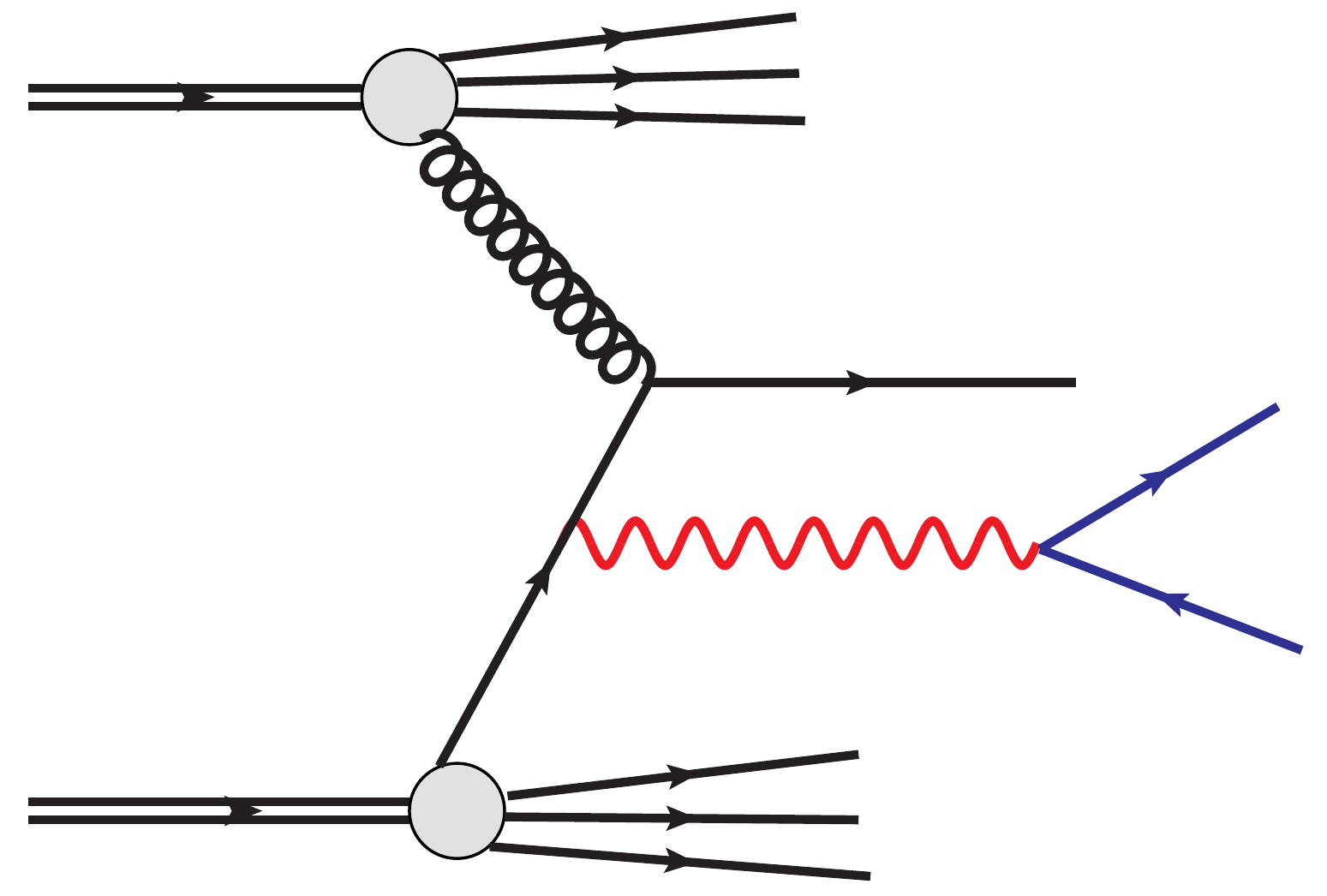}
\includegraphics[width=.5\textwidth]{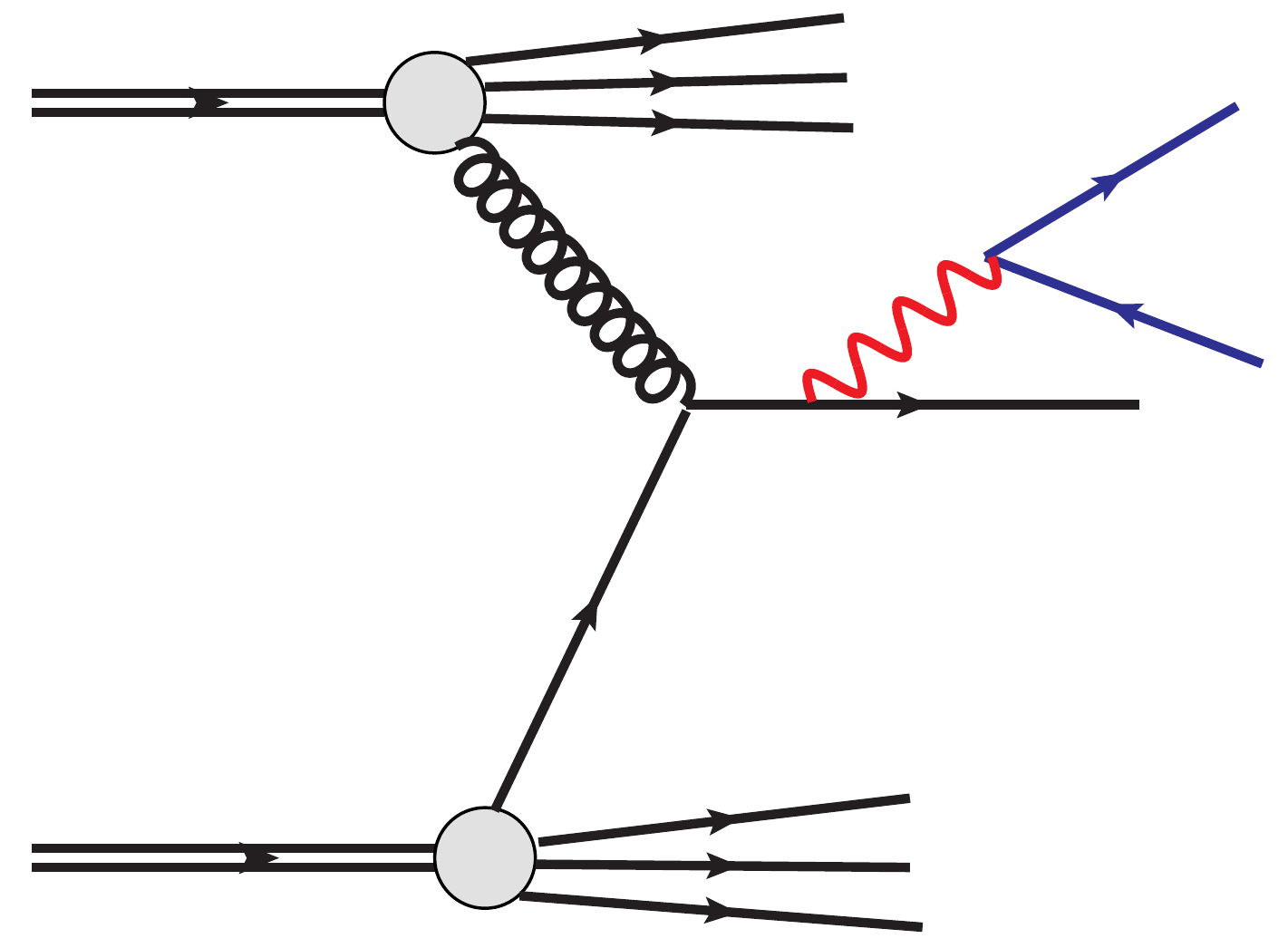}
\caption{\label{fig:diagrams}
The diagrams relevant for forward and backward production of dilepton pairs.
}
\end{figure}

The mechansims considered are shown in the diagrams in Fig.\ref{fig:diagrams}.

\begin{figure}[!ht]
\includegraphics[width=.5\textwidth]{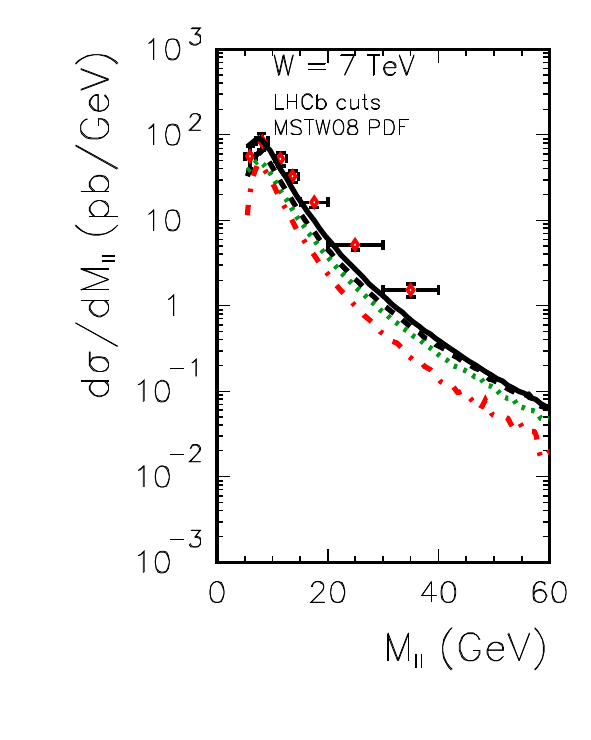}
\includegraphics[width=.5\textwidth]{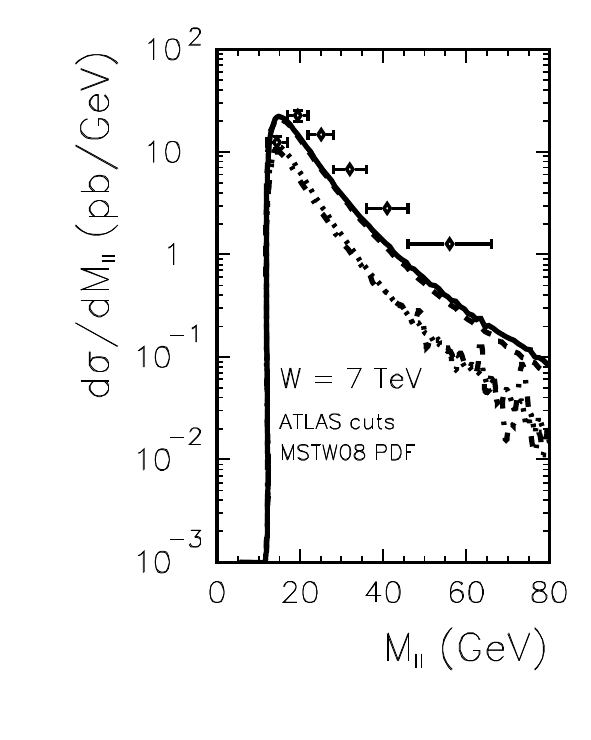}
  \caption{\label{fig:lhcb_dsig_dMll_ugdfs}
Left panel: Invariant mass distribution (only the dominant component)
for the LHCb cuts: 2 $< y_{+},y_{-} <$ 4.5, $k_{T+},k_{T-} >$ 3 GeV
for different UGDFs: KMR (solid), Kutak-Stasto (dashed), AAMS (dotted)
and GBW (dash-dotted). Right panel:  the same for the ATLAS kinematics:
-2.4 $< y_{+},y_{-} <$ 2.4, $k_{T+}, k_{T-} >$ 6 GeV.
Here both $g q/\bar q$ and $q/\bar q g$ contributions have been included.
}
\end{figure}

\begin{figure}[!ht]
\includegraphics[width=6cm]{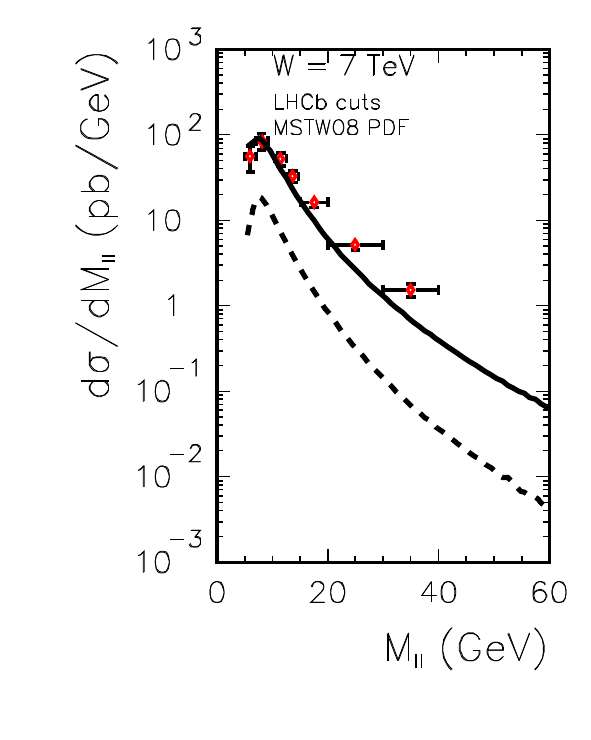}
\includegraphics[width=6cm]{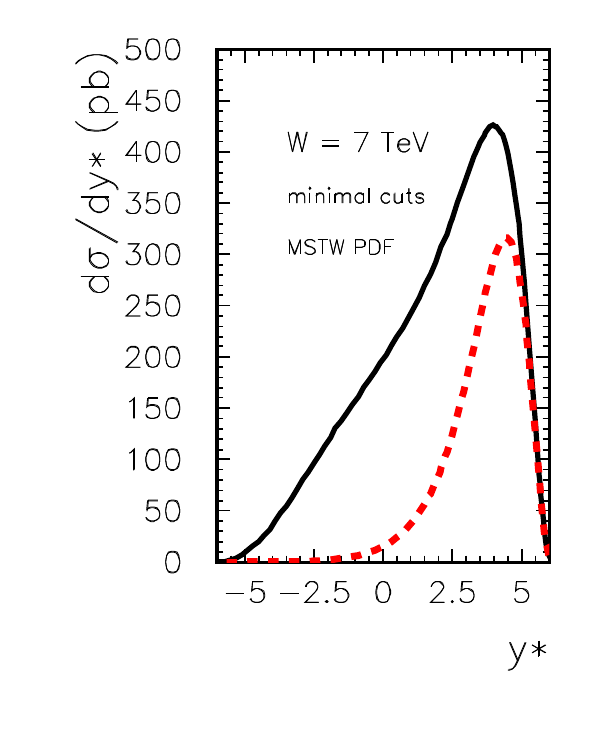}
  \caption{\label{fig:lhcb_dsig_dMll_second}
Left Panel: By the dashed line, we show the contributions of the second-side 
component for the LHCb kinematics: 
2 $< y_{+},y_{-} <$ 4.5, $k_{T+},k_{T-} >$ 3 GeV. KMR UGDF was used here.
Right panel: Distribution in rapidity of the dileptons
for $\sqrt{s}$ = 7 TeV and $k_{T+},k_{T-} >$ 3 GeV
for MSTW08 valence quark distributions and KMR UGDFs.
The dashed line is the contribution from valence quarks only.
}
\end{figure}
\section{Results}
We start by defining the relevant kinematical variables.
Below, $x_\pm$ will denote the 
longitudinal (lightcone-) momentum fractions of leptons, while $\bk_\pm$ are their transverse momenta.
The heavy virtual photon of mass $M^2$ then carries the longitudinal momentum fraction $x_F = x_+ + x_-$ and 
transverse momentum $\bq = \bk_+ + \bk_-$.
It is useful to introduce also the light-front relative transverse momentum of $l^+$ and $l^-$:
\begin{eqnarray}
 \bl = {x_+ \over x_F} \bk_- - {x_- \over x_F} \bk_+ \, .
\end{eqnarray}
Then, the inclusive cross section for lepton pair production can be written in the form:
\begin{eqnarray}
{d\sigma(pp \to l^+ l^- X) \over dx_+ dx_- d^2\bk_+ d^2\bk_-} 
&=&{ \aem \over (2 \pi)^2 M^2} {x_F \over x_+ x_-} \Big\{ \Sigma_T(x_F,\bq,M^2)  D_T\Big({x_+ \over x_F} \Big)  
\nonumber \\
&+&
\Sigma_L(x_F,\bq,M^2)  D_L\Big({x_+ \over x_F}\Big) \nonumber \\
&+& \Sigma_\Delta(x_F,\bq,M^2)  D_\Delta\Big({x_+ \over x_F}\Big)  
\Big({\bl \over |\bl|} \cdot {\bq \over |\bq|} \Big) \nonumber \\
&+&  \Sigma_{\Delta \Delta}(x_F,\bq,M^2) D_{\Delta \Delta}\Big({x_+
  \over x_F}\Big) \Big(2  \Big({\bl \over |\bl|} \cdot {\bq \over |\bq|}
\Big)^2 - 1 \Big) \Big\} . 
\end{eqnarray}

The functions $\Sigma_i(x_F,\bq,M^2),\, i = T,L,\Delta,\Delta\Delta$ are 
in a one-to-one correspondence with the four helicity structure functions 
\cite{Oakes:1966} of inclusive lepton pair production in a Gottfried-Jackson frame. 
They contain all information of strong dynamics in the production
of the virtual photon. The functions $D_i$ and the momentum structures in brackets 
represent the density matrix of decay of the massive photon into $l^+ l^-$. For explicit expressions, see \cite{Schafer:2016qmk}.

Let us concentrate now on one of the partonic subprocesses, where
a fast quark from one proton radiates a virtual photon while interacting with a small-$x$ gluon of the other proton.
(E.g. the top two diagrams in Fig. \ref{fig:diagrams}.)
Naturally the large-$x$ quark is described by the collinear quark distribution, while for the
low-$x$ gluon it is more appropriate to use the $k_T$-dependent unintegrated gluon distribution.

We can then write for the functions $\Sigma_i$ an impact-factor representation typical
of the $k_\perp$-factorization:
\begin{eqnarray}
\Sigma_i (x_F,\bq,M) 
&=& \sum_f  {e_f^2 \alpha_{\rm em} \over 2 N_c} \int_{x_F}^1  dx_1 \,
\Big[q_f(x_1,\mu^2) + \bar q_f(x_1,\mu^2 )\Big]  
\nonumber \\
&\times&
\int {d^2 \bkappa \over \pi \bkappa^4} 
 {\cal{F}}(x_2,\bkappa^2) 
\alpha_S(\bar q^2) I_i \Big( {x_F \over x_1} ,\bq,\bkappa \Big) \, . 
\end{eqnarray}
Here appears the unintegrated gluon distribution 
\begin{eqnarray}
 {\cal{F}}(x_2,\bkappa^2) \propto {\partial x_2 g(x_2, \bkappa^2) \over \partial \log(\bkappa^2)} \, .
\end{eqnarray}
The impact factors $I_i$ can be found in \cite{Schafer:2016qmk}.

Here an important comment on the longitudinal momentum fractions $x_1, x_2$ is in order.
They must be obtained from the full $l^+ l^- q$ final state:
\begin{eqnarray}
x_1 &=& \sqrt{ \bk_+^2  \over S} e^{y_+} + \sqrt{ \bk_-^2 \over S} e^{y_-} + 
{\sqrt{ \bk_q^2 \over S} e^{y_q}} \; , \nonumber \\
x_2 &=& \sqrt{ \bk_+^2 \over S} e^{-y_+} + \sqrt{ \bk_-^2  \over S} e^{-y_-} + 
{\sqrt{ \bk_q^2  \over S} e^{-y_q}} \; .
\label{x1_x2}
\end{eqnarray}
Neglecting the contribution from the final state (anti-)quark leads to a systematic 
underestimation of $x$-values, which may artificially enhance saturation effects.

In Fig. \ref{fig:lhcb_dsig_dMll_ugdfs} we compare our results to recent experimental data.
In the left panel we compare our results for the dilepton invariant mass distribution 
to the data from the LHCb collaboration \cite{LHCb-CONF_2012},
which cover the forward rapidity region. Here a reasonable description of
data can be obtained by an unintegrated gluon distribution constructed by the KMR prescription.
Other gluon distributions which include gluon saturation effects do not lead to such a good
agreement.
In the right panel, we compare our results to the ATLAS data \cite{Aad:2014qja}.
These data were obtained in the central rapidity region. This kinematical domain
is strictly speaking beyond the region of applicability of our approach.
The asymmetrical treatment of collinear quarks and $k_T$-dependent gluons is not
warranted here. And indeed, we do not describe the ATLAS data well, especially at large invariant
masses.

The results shown in Fig. \ref{fig:lhcb_dsig_dMll_second} were obtained in the LHCb 
kinematics. In the left panel we show
by the dashed line the contribution from dileptons emitted from the ``other side'' proton. 
As we observe, such a spillover of dileptons emitted into the forward 
region of ``the other'' proton is not negligible. 
It seems to have been generally neglected in dipole model calculations. 
In the right panel we show the rapidity distribution of the virtual photon. 
By the red dashed line we show the contribution from valence quarks of the 
``forward'' proton only. We see that within the rapidity coverage of LHCb
sea quarks are important.

\section{Summary}
In this talk at the Low-$x$ meeting, we have presented the main results from our
recent paper \cite{Schafer:2016qmk} on the Drell-Yan production of
dileptons in the forward rapidity region in a hybrid factorization
approach. Here the large-$x$ parton participating in the hard process
is described by a collinear parton distribution, while for the low-$x$ parton 
an unintegrated parton distribution is taken.

We have compared the results of our calculations 
to recent experimental data for low-mass dilepton production 
from the LHCb and ATLAS experiments.

Going beyond on previous work in the literature, we have found
that emissions from both protons have to be included even for
the LHCb configuration.

We find that LHCb data do not require gluon saturation effects 
at small $M_{ll}$.

{\bf{Acknowledgements:}}\\
The work reported here was supported by the Polish National Science 
Centre grant DEC-2014/15/B/ST2/02528.


\begin{thebibliography}{10}


\bibitem{Gelis:2002fw} 
  F.~Gelis and J.~Jalilian-Marian,
  Phys.\ Rev.\ D {\bf 66}, 094014 (2002)
  [hep-ph/0208141];
  Phys.\ Rev.\ D {\bf 76}, 074015 (2007)
  [hep-ph/0609066].

\bibitem{Ducati:2013cga} 
  M.~B.~G.~Ducati, M.~T.~Griep and M.~V.~T.~Machado,
  Phys.\ Rev.\ D {\bf 89}, no. 3, 034022 (2014)
  [arXiv:1307.6882 [hep-ph]].

\bibitem{GolecBiernat:2010de} 
  K.~Golec-Biernat, E.~Lewandowska and A.~M.~Stasto,
  Phys.\ Rev.\ D {\bf 82}, 094010 (2010)
  [arXiv:1008.2652 [hep-ph]].

\bibitem{Basso:2015pba} 
  E.~Basso, V.~P.~Goncalves, J.~Nemchik, R.~Pasechnik and M.~Sumbera,
  Phys.\ Rev.\ D {\bf 93}, 034023 (2016)
  [arXiv:1510.00650 [hep-ph]].

\bibitem{Motyka:2014lya} 
  L.~Motyka, M.~Sadzikowski and T.~Stebel,
  JHEP {\bf 1505}, 087 (2015)
  [arXiv:1412.4675 [hep-ph]].

\bibitem{Schafer:2016qmk}
  W.~Sch\"afer and A.~Szczurek,
  Phys.\ Rev.\ D {\bf 93} (2016) no.7,  074014
  doi:10.1103/PhysRevD.93.074014
  [arXiv:1602.06740 [hep-ph]].

\bibitem{Oakes:1966}
R.~J.~Oakes, Nuovo\ Cimento\ A {\bf 44}, 440 (1966);
  C.~S.~Lam and W.~K.~Tung,
  Phys.\ Rev.\ D {\bf 18}, 2447 (1978).


\bibitem{LHCb-CONF_2012} 
  [The LHCb collaboration],
  ``Inclusive low mass Drell-Yan production in the forward region at sqrt(s)=7 TeV ,''
  LHCb-CONF-2012-013; Conference report prepared for XX International Workshop on Deep-Inelastic Scattering and Related Subjects, 26-30, March 2012, Bonn, Germany.
  
  
\bibitem{Aad:2014qja} 
  G.~Aad {\it et al.} [ATLAS Collaboration],
  JHEP {\bf 1406}, 112 (2014)
  [arXiv:1404.1212 [hep-ex]].





\end{thebibliography}
\end{document}